\documentclass[superscriptaddress,secnumarabic,amssymb,amsmath,nobibnotes,aps,prd,showkeys,nofootinbib,preprint]{revtex4}

\usepackage{graphicx}
\usepackage{float}
\usepackage{bm}
\usepackage{amsmath}
\usepackage{amsfonts}
\usepackage{amssymb}
\usepackage{epstopdf}
\usepackage{natbib}%
\newcommand{\bee}{\begin{equation}}
\newcommand{\eee}{\end{equation}}
\newcommand{\eaa}{\end{eqnarray}}
\newcommand{\baa}{\begin{eqnarray}}

\usepackage{color}

\begin{document}

\title{Gauge Symmetry of the Chiral Schwinger model from an improved Gauge Unfixing formalism}

\author{Gabriella V. Ambrósio}
\email{gabriellambrosio@gmail.com}
\affiliation{Department of Physics, Federal University of Juiz de Fora , 36036-330, Juiz de Fora, MG, Brazil}
\author{Cleber N. Costa}
\email{cleber.costa@ice.ufjf.br}
\affiliation{Department of Physics, Federal University of Juiz de Fora , 36036-330, Juiz de Fora, MG, Brazil}
\author{Paulo R. F. Alves}
\email{paulo.alves@ice.ufjf.br}
\affiliation{Department of Physics, Federal University of Juiz de Fora , 36036-330, Juiz de Fora, MG, Brazil}
\author{Everton M. C. Abreu}
\email{everton.abreu.gmail.com}
\affiliation{Department of Physics, Federal University of Rural do Rio de Janeiro, 23890-971, Seropédica, RJ, Brazil}
\author{Jorge Ananias Neto}
\email{jorge@fisica.ufjf.br}
\affiliation{Department of Physics, Federal University of Juiz de Fora , 36036-330, Juiz de Fora, MG, Brazil}
\author{Ronaldo Thibes}
\email{thibes@uesb.edu.br}
\affiliation{Department of Exact and Natural Sciences,
State University of Sudoeste da Bahia,
Rodovia BR 415, Km 03, S/N, Itapetinga – 45700-000, Brazil}

\begin{abstract}
In this paper, the Hamiltonian structure of the bosonized chiral Schwinger model (BCSM) is analyzed. From the consistency condition of the constraints obtained from the Dirac method, we can observe that this model presents, for certain values of the $\alpha$ parameter, two second-class constraints, which means that this system does not possess gauge invariance. However, we know that it is possible to disclose gauge symmetries in such a system by converting the original second-class system into a first-class one. This procedure can be done through the gauge unfixing (GU) formalism by acting with a projection operator directly on the original second-class Hamiltonian, without adding any extra degrees of freedom in the phase space. One of the constraints becomes the gauge symmetry generator of the theory and the other one is disregarded. At the end, we have a first-class Hamiltonian satisfying a first-class algebra. Here, our goal is to apply a new scheme of embedding second-class constrained systems based on the GU formalism, named improved GU formalism, in the BCSM. The original second-class variables are directly converted into gauge invariant variables, called GU variables. We have verified that the Poisson brackets involving the GU variables are equal to the Dirac brackets between the original second-class variables. Finally, we have
found that our improved GU variables coincide with those obtained from an improved BFT method after a particular choice for the Wess-Zumino terms.
\end{abstract}

\maketitle

\section{Introduction}\label{sec:Introduction} 
It has been shown over the last decades that anomalous gauge theories in two dimensions can be consistently and unitarily quantized for both Abelian~\cite{A_3, A_4} and non-Abelian~\cite{A_5, A_6} cases. In this scenario, the two dimensional model that has been extensively studied is the Chiral Schwinger model. This well known anomalous gauge theory~\cite{A_4, A_7, A_8} involves chiral fermions coupled to a U(1) gauge field in $1+1$ dimensions and is defined by

\begin{align}
\mathcal{L}=-\frac{1}{4}F_{\mu \nu}F^{\mu \nu}+\bar{\psi}\left [ i\gamma^{\mu} \partial_{\mu} + e\sqrt{\pi}\gamma^{\mu}A_{\mu}(1+ \gamma_{5}) \right ]\psi \,. \label{z1}
\end{align}
Classically, the theory has gauge invariance, but this is lost upon quantization~\cite{A_2}. We will pay attention to its bosonized version, which gauge invariance is already lost in its corresponding classical theory itself. This model has attracted much attention over the last decade, mainly in the context of string theories~\cite{A_17}, and also due to the huge progress in understanding the physical meaning of anomalies in quantum field theories~\cite{A_18}. This theory is represented as

\begin{align}
\mathcal{L}= -\frac{1}{4}F_{\mu \nu}F^{\mu \nu}+\frac{1}{2}(\partial_{\mu} \varphi )^{2}+ e(g^{\mu \nu}- \epsilon ^{\mu \nu})(\partial_{\mu} \varphi)A_{\nu}+\frac{1}{2}e^{2}\alpha A_{\mu}^{2} \,. \label{z2}
\end{align}
This version has been carefully analyzed and found to yield a consistent theory for a class of regularizations~\cite{A_4}. The class of regularization that have been considered involves a dimensionless parameter $\alpha$~\cite{A_9}. This has to satisfy $\alpha \geqslant 1$ for the theory to be consistent. For $\alpha > 1$, the system is seen to have two second-class constraints. For $\alpha = 1$, there are four constraints, where two of them are first-class. For $\alpha < 1$, the theory is non-unitary~\cite{A_8}. The theory is gauge non-invariant for all values of $\alpha$. In this paper, we will consider only the case $\alpha > 1$.

We can find several works dealing with obtaining gauge invariant theories. Gauge invariant theories are extremely important because for a theory to be
quantized it must be a gauge theory, and quantization of second-class system is much more difficult when compared to first-class theories~\cite{mes}.
Here we highlight the work of Faddeev and Shatashvili~\cite{A_10} which introduced Wess-Zumino terms into the effective action of a theory and is responsible for transforming second-class systems into first-class ones. This method is known as Batalin-Fradkin-Tyutin (BFT) formalism~\cite{A_11, A_12}. 
Since then there has been diverse applications of the BFT formalism \cite{bft1,bft2,bft3,bft4,bft5}.
On the other hand, Vytheeswaran~\cite{A_8, A_15, A_16},  based on the idea of Mitra and Rajaraman~\cite{A_13}, that does not make use of extra phase-space variables, applied the Lie projection operator directly in
the second-class constrained Hamiltonians in order to reveal hidden symmetries. 
Then, the motivation of this paper is to apply an improved GU formalism~\cite{A_14,proto,A_20} in the Hamiltonian structure of the BCSM to obtain a gauge invariant theory. The improved GU formalism consists of redefining the phase space variables by making them first-class. This feature of making modifications directly to the phase space variables is what makes this method advantageous over others.

We have organized this paper as follows: in Section 2 we have introduced the improved GU formalism. In Section 3 we have examined the BCSM with the use of the Dirac constrained method. In Section 4 the improved GU formalism is applied to the BCSM. In Section 5 the results of the improved GU 
are compared with those of the improved BFT method. In Section 6 we have given our final considerations.

\section{The improved gauge unfixing formalism}
\label{sqm}
The improved GU formalism developed by Neto~\cite{A_18,A_14,proto} is based on the initial idea of the GU formalism~\cite{A_13, A_15, A_16} which is to convert the second-class Hamiltonian into first-class one. Here we consider any second-class function of the phase space, namely, $G(A_{\mu},\pi_{\mu})$. Thus, our strategy is to write a first-class function $\tilde{G}(\tilde{A}_{\mu},\tilde{\pi}_{\mu})$ from the second-class function $G$ by redefining the original phase space variables. At the end, the function $\tilde{G}$ need to satisfy the following variational condition

\begin{align}
  \delta \tilde{G} = \varepsilon \left \{ \tilde{G} , \chi \right \}=0 \,, \label{3.8}
\end{align}
where $\varepsilon$ is an infinitesimal parameter and $\chi$ is a scaled second-class constraint chosen to be the gauge symmetry generator. 
Here we are considering that the system has only two second-class constraints which are denoted by $Q_1$ and $Q_2$. Any function of $\tilde{G}$ will be gauge invariant since

\begin{align}
\left \{ \tilde{G}, \chi \right \}= \left \{ \tilde{A}, \chi \right \} \frac{\partial  \tilde{G} }{\partial \tilde{A}} + \frac{\partial \tilde{G}}{\partial \tilde{\pi }}\left \{ \tilde{\pi }, \chi \right \} = 0 \,.
\end{align}
Consequently, we can obtain a gauge invariant function from the replacement of

\begin{align}
G(A_{i}, \pi_{i}) \rightarrow G(\tilde{A_{i}}, \tilde{ \pi_{i}})=\tilde{G}(\tilde{A_{i}}, \tilde{ \pi_{i}}) \,.
\end{align}
The gauge invariant phase space variables $\tilde{G}$ are constructed by the series in powers of the discarded constraint $Q_{2}$

\begin{align}
\tilde{G}(x)=G(x)+\int dyb_{1}(x,y)Q_{2}(y)+\iint dydz b_{2}(x,y,z)Q_{2}(y)Q_{2}(z)+... \,, \label{3.11}
\end{align}
which has the following boundary condition on the constraint surface $Q_{2}=0$

\begin{align}
    \tilde{G}_{(Q_{2}=0)}=G \,,
\end{align}
that is, we need to recover the original second-class system when $Q_2=0$. The coefficients $b_{n}$ in the relation~\eqref{3.11} are then determined by the variational conditional, Eq.~\eqref{3.8}. The general equation for $b_{n}$ is

\begin{align}
\delta \tilde{G}(x)=\delta G(x)+\delta \int dyb_{1}(x,y)Q_{2}(y)+ \delta \iint dydz b_{2}(x,y,z)Q_{2}(y)Q_{2}(z)+...=0 \,, \label{3.18}
\end{align}
where

\begin{align}
 & \delta G(x)=\int dy \varepsilon (y)\left \{ G(x), \chi(y) \right \}, \\
 &\delta b_{1}(x)=\int dy \varepsilon (y)\left \{ b_{1}(x), \chi(y) \right \},\\
&\delta Q_{2}(x)=\int dy \varepsilon (y)\left \{ Q_{2}(x), \chi(y) \right \} = - \varepsilon (x) \,. \label{3.17}
\end{align}
So, for the first order correction term $(n = 1)$ we have from Eq.~\eqref{3.18} 

\begin{align}
    \delta G(x)+\int dyb_{1}(x,y)\delta Q_{2}(y)=0 \qquad \Rightarrow \qquad b_{1}(x,y)=\frac{\delta G(x)}{\varepsilon (x)} \delta(x-y) \,. \label{3.19}
\end{align}
For the second order correction term $(n = 2)$ we obtain

\begin{eqnarray}
    \int dy\delta b_{1}(x,y)Q_{2}(y)+2\iint dydz b_{2} (x,y,z)\delta Q_{2}(y)Q_{2}(z)=0  \nonumber \\
    \Rightarrow  b_{2}(x,y,z)=\frac{\delta \delta G(x)}{2\varepsilon ^{2}(x)}\delta (x-y)\delta(y-z) \,. \label{3.20}
\end{eqnarray}
For $n \geq 2$, the general relation is 

\begin{align}
    b_n(x_1,x_2,x_3,...,x_n) = \frac{\delta^{n}G(x_1)}{n!\varepsilon^{n}(x_1)} \delta(x_1-x_2) \delta(x_2-x_3).......\delta(x_{n-1}-x_n) \,.
\end{align}
Once the coefficients $b_n$ are all determined, we can substitute them in Eq.~\eqref{3.11}, resulting in the expression for the gauge invariant function $\tilde{G}(x)$ given by

\begin{eqnarray}
\label{gt} 
    \tilde{G}(x)= \left ( 1 +Q_{2}(x)\frac{\delta}{\varepsilon (x)}+ \frac{1}{2!}Q_{2}^{2}(x)\frac{\delta \delta}{\varepsilon ^{2}(x)}+... \right )G(x) \,. \label{3.21}
\end{eqnarray}
Eq.~(\ref{gt}) can also be written in terms of a projection operator on $G$ in the following form

\begin{eqnarray}
\label{proj}
\tilde{G}(x) =  e^{Q_{2}(x)\frac{\delta}{\varepsilon }}:G(x) \,,
\end{eqnarray}
where an ordering prescription must be adopted: $Q_{2}$ must come before $\frac{\delta}{\varepsilon }$. 
Therefore, from the GU variables defined in Eq.~(\ref{gt}) or (\ref{proj}) we can derive a corresponding gauge invariant theory.

\section{The bosonized Chiral Schwinger Model}

The BCSM, being a (1 + 1) dimensional field theory, is described by the following Lagrangian density

\begin{eqnarray}
\label{lagrangian}
\mathcal{L}= -\frac{1}{4}F_{\mu \nu}F^{\mu \nu}+\frac{1}{2}(\partial_{\mu} \varphi )^{2}+ e(g^{\mu \nu}- \epsilon ^{\mu \nu})(\partial_{\mu} \varphi)A_{\nu}+\frac{1}{2}e^{2}\alpha A_{\mu}^{2}\,, \label{a}
\end{eqnarray}
where $F_{\mu \nu} = \partial_{\mu}A_{\nu} - \partial_{\nu}A_{\mu}$ is the electromagnetic field tensor, $g_{\mu \nu} = (+1,-1)$, $\epsilon 
^{01} = -\epsilon^{10}= 1$ and $\alpha$ is the regularisation parameter. From the Lagrangian density (\ref{lagrangian})  we can obtain the canonical momenta which are given by

\begin{equation}
   \pi_{0} = -F_{00} = 0 \,, \label{b} 
\end{equation}
\begin{equation}
\pi_{1} = -F_{01} = -(\partial_{0} A_{1} - \partial_{1}A_{0}) = -\dot{A}_{1}+\partial_{1}A_{0} \,, 
\end{equation}
\begin{equation}
    \pi_{\varphi} = \frac{\partial \mathcal{L}}{\partial(\partial_{0}\varphi )} = \partial_{0}\varphi + e(A_{0}-A_{1}) \,. \label{4.6}
\end{equation}
From Eq.~\eqref{b} we have the primary constraint

\begin{eqnarray}
\label{omega1}
\Omega_{1} \equiv \pi_{0} \approx  0 \,. \label{c}
\end{eqnarray}
Using the Legendre transformation

\begin{eqnarray}
H_{c} = \int dx \left ( \pi^{\mu}\dot{A}_{\mu} + \pi_{\varphi}\dot{\varphi} - \mathcal{L} \right ) \,
\end{eqnarray}
we can obtain the canonical Hamiltonian

\begin{align}
H_{c} = \int dx &\bigg[ \frac{1}{2} \pi_{1}^{2} + \frac{1}{2} \pi_{\varphi }^{2} + \frac{1}{2}(\partial_{1} \varphi)^{2} + e\big(\partial_{1}\varphi + \pi_{\varphi}\big)A_{1} + \frac{1}{2}e^{2} \big(\alpha + 1\big)A_{1}^{2}\nonumber\\ 
& - A_{0} \big [ -\partial_{1} \pi_{1} + \frac{e^{2}(\alpha - 1)}{2}A_{0} + e(\partial_{1}\varphi + \pi_{\varphi}) + e^{2}A_{1} \big ]\bigg ] \,. \label{6}
\end{align}
By writing the primary Hamiltonian as 

\begin{eqnarray}
H_{p} = H_{c} + \int dx \lambda_{1} \Omega_{1} \,,
\end{eqnarray}
where $\lambda_1$ is the Lagrange multiplier, and demanding the time stability condition of the primary constraint, Eq.~\eqref{c}, we have the secondary constraint

\begin{eqnarray}
\label{sec1}
 \Omega_{2} \equiv -\partial _{1}\pi_{1} + e^{2}(\alpha-1)A_{0} + e(\partial_{1}\varphi + \pi_{\varphi}) + e^{2}A_{1} \approx 0 \,, \label{d}
\end{eqnarray}
which is identified as being a modified Gauss law. We note that no more constraints are produced by this iterative process. Therefore $\Omega_{1}$ and $\Omega_{2}$ are the total constraints of
the model. Calculating the Poisson brackets of these constraints, we have

\begin{align}
\label{alfa}
\varepsilon _{12} = \left\{\Omega_{1}(x), \Omega_{2}(y) \right\} = -e^{2}(\alpha - 1)\delta(x-y) \,.
\end{align}
From Eq.~(\ref{alfa}) we can see that for $\alpha = 1$ the constraints $\Omega_1$ and $\Omega_2$ satisfy a first-class algebra and for $\alpha > 1$ the constraints 
$\Omega_1$ and $\Omega_2$ satisfy a second-class algebra. We shall only consider 
the case $\alpha > 1$ where the BCSM has two second-class constraints.

\section{Discussion of results}
In this section, we will reveal the hidden symmetries of the BCSM with the use of the improved GU formalism. As we have seen, the BCSM presents two second-class constraints for $\alpha > 1$ which are  Eqs.~\eqref{c} and~\eqref{d}. Then, we can choose one of the these two second-class constraints to form the gauge symmetry generator and the other one will be discarded. Thus, we have two possible choices for the gauge symmetry generator for the BCSM. We will consider these two different cases separately:

\subsection*{Case I}

In this case we will initially rescale the primary constraint $\Omega_1=\pi_0$ to build a gauge symmetry generator $ \psi$

\begin{eqnarray}
\label{g1s}
\psi \equiv -\frac{\Omega_{1}}{e^{2}(\alpha -1)}= -\frac{\pi_{0}}{e^{2}(\alpha -1)} \,,
\end{eqnarray}
so that $\left \{ \psi(x), \Omega_{2}(y) \right \}=\delta(x-y)$. The second-class constraint $\Omega_{2}$, Eq.~(\ref{sec1}), will be discarded and the gauge invariant quantities will be generated only by $\psi$, Eq.~(\ref{g1s}). The infinitesimal gauge transformations generated by $\psi$ are

\begin{align}
\delta A_{0}(x)&=\int dy \varepsilon (y)\left \{ A_{0}(x), \psi(y) \right \}=-\frac{\varepsilon (x)}{e^{2}(\alpha-1)},\label{e} \\
\delta \pi_{1}(x)&=\delta A_{1}(x)=\delta \varphi(x)= \delta \pi_{\varphi}(x)=0 \,,\\
\delta \Omega_{2}(x)&=\int dy \varepsilon (y)\left \{ \Omega_{2}(x), \psi(y) \right \} 
=-\varepsilon (x) \,.\label{f} 
\end{align}
Then, we can see that only $A_{0}$ is not gauge invariant. So, the gauge invariant variable $\Tilde{A}_{0}$ is constructed by the power series in $\Omega_{2}$
\begin{eqnarray}
\label{a0t}
\tilde{A}_{0}(x)= A_{0}(x)+\int dy b_{1}(x,y)\Omega_{2}(y)+\iint dy dz b_{2}(x,y,z)\Omega_{2}(y)\Omega_{2}(z)+... \,.
\end{eqnarray}
Applying the variational condition $\delta \tilde{A}_{0}=0$, we can compute all the correction terms $b_{n}$. For the linear correction term in order of $\Omega_{2}$ we obtain

\begin{eqnarray}
\label{da0}
\delta A_{0}(x)+\int dy  b_{1}(x,y)\delta \Omega_{2}(y)=0 \,. \label{g}
\end{eqnarray}
Combining Eqs.~\eqref{e}, \eqref{f} and (\ref{da0}) we have

\begin{eqnarray}
b_{1}(x,y)= -\frac{1}{e^{2}(\alpha  -1)}\delta(x-y) \,. \label{h}
\end{eqnarray}
For the quadratic term we have $b_2=0$. Then, all the correction terms $b_{n}$ are null for $n\geq 2$. Taking this into consideration and using Eq.~\eqref{h} in (\ref{a0t}), we can derive the GU variable $\tilde{A}_{0}$ 

\begin{equation}
\label{a0t1}
\tilde{A}_{0}(x)=A_{0}(x)-\frac{1}{e^{2}(\alpha -1)}\Omega_{2}(x) \,.
\end{equation}
For the choice of gauge symmetry generator, Eq.~(\ref{g1s}), the fields $A_{1}$, $\pi_{1}$, $\varphi$ and $\pi_{\varphi}$ are already gauge invariant. Therefore, the others GU variables are 

\begin{eqnarray}
&\tilde{A}_{1}= A_{1}\,,\\
&\tilde{\pi}_{1}=\pi_{1},\\
&\tilde{\varphi}=\varphi\,,\\
&\tilde{\pi}_{\varphi}=\pi_{\varphi}\,.
\end{eqnarray}
Replacing the second-class variables $A_{0}, A_{1}, \varphi, \pi_{1}$ and $\pi_{\varphi}$ by the GU variables $\tilde{A}_{0},\tilde{A}_{1},\tilde{\varphi},\tilde{\pi}_{1}$and $\tilde{\pi}_{\varphi}$ in the canonical Hamiltonian, Eq.~\eqref{6}, thus we can obtain the gauge invariant Hamiltonian $\tilde{H}_{c}$ 

\begin{align}
\tilde{H}_c&= H_c + \int dx \, \frac{1}{2 e^2 (\alpha -1)} \, \Omega_2^2 \,,
\end{align}
which by construction  satisfies $\left \{ \psi, \Tilde{H}_{c} \right \}=0$.

The Poisson brackets between the GU variables are

\begin{align}
 \left \{ \tilde{A}_{0}(x),\tilde{\pi}_{1}(y) \right \}&=\frac{1}{(\alpha - 1)}\delta(x-y) = \left \{ A_{0}(x),\pi_{1}(y) \right \}_{D},\\
\left \{ \tilde{A}_{0}(x),\tilde{\varphi}(y) \right \}&=\frac{1}{e(\alpha - 1)} \delta(x-y)=\left \{ A_{0}(x),\varphi(y) \right \}_{D},   
\end{align}
\begin{align}
\left \{ \tilde{A}_{0}(x),\tilde{\pi}_{\varphi}(y) \right \}&=-\frac{1}{e(\alpha -1)}\partial_{1}\delta(x-y)=\left \{ A_{0}(x),\pi_{\varphi}(y) \right \}_{D},\\
\left \{ \tilde{A}_{0}(x),\tilde{A}_{1}(y) \right \}&= \frac{1}{e^{2}(\alpha -1)}\partial_{1}\delta(x-y) = \left \{ A_{0}(x),A_{1}(y) \right \}_{D},\\
\left \{ \tilde{A}_{1}(x),\tilde{\pi}_{1}(y) \right \}&=-\delta(x-y) =\left \{ A_{1}(x),\pi_{1}(y) \right \}_{D},\\
\left \{ \tilde{\varphi}(x),\tilde{\pi}_{\varphi}(y) \right \}&= \delta(x-y) =\left \{ \varphi(x),\pi_{\varphi}(y) \right \}_{D}.
\end{align}
We can observe that the Poisson brackets calculated between the GU variables agree with the results obtained by using the Dirac brackets calculated between the original phase space variables. These above equalities between the Poisson and Dirac brackets can be demonstrated in a general way in refs.~\cite{db,A_22}.

\subsection*{CASE II}

In this case we initially will rescale the secondary constraint $\Omega_{2}$, Eq. (\ref{sec1}), with the objective of building a gauge symmetry generator $\chi$

\begin{eqnarray}
\label{rsecond}
\chi \equiv \frac{1}{e^{2}(\alpha - 1)}\left [ -\partial _{1}\pi_{1} + e^{2}(\alpha-1)A_{0} + e(\partial_{1}\varphi + \pi_{\varphi}) + e^{2}A_{1}
\right ] \,,
\end{eqnarray}
such that $\left \{ \chi (x),\Omega_{1}(y) \right \}=\delta(x-y)$. The primary constraint, $\Omega_{1}=\pi_0$, will be discarded.
The infinitesimal gauge transformations generated by $\chi$ are

 \begin{align}
&\delta A_{0}(x)=\int dy \varepsilon (y)\left \{ A_{0}(x), \chi(y) \right \}=0\,, \label{4.66}\\
&\delta A_{1}(x)=\int dy \varepsilon (y)\left \{ A_{1}(x), \chi(y) \right \}=-\frac{1}{e^{2}(\alpha -1)}\,\partial_{1}\varepsilon (x)\,, \label{4.67}\\
&\delta \varphi(x)=\int dy \varepsilon (y)\left \{ \varphi(x), \chi(y) \right \}=\frac{1}{e(\alpha -1)}\,\varepsilon (x)\,,\label{4.68}\\
&\delta \pi_{\varphi}=\int dy \varepsilon (y)\left \{ \pi_{\varphi}(x), \chi(y) \right \}=\frac{1}{e(\alpha -1)}\,\partial_{1}\varepsilon (x)\,,\label{4.69}\\
&\delta \pi_{1}(x)=\int dy \varepsilon (y)\left \{ \pi_{1}(x), \chi(y) \right \}=\frac{1}{(\alpha -1)}\,\varepsilon (x)\,, \label{4.70}\\
&\delta \Omega_{1}(x)=\frac{1}{e^{2}(\alpha -1)}\int dy \varepsilon (y)\left \{ \Omega_{1}(x), \chi(y) \right \}=-\varepsilon (x)\,. \label{4.71}
\end{align}
For this choice we can see that only $A_{0}$ is already a first-class variable. Thus, the GU quantities $\tilde{A}_{1}$, $\tilde{\varphi}$, $\tilde{\pi}_{1}$ and $\tilde{\pi}_{ \varphi}$ must be derived. For the gauge invariant variable $\tilde{A}_{1}$ we have the following series
in $\Omega_1$

\begin{align}
\tilde{A}_{1}(x)= A_{1}(x)+\int dy c_{1}(x,y)\Omega_{1}(y)+\iint dy dz c_{2}(x,y,z)\Omega_{1}(y)\Omega_{1}(z)+... \,. \label{37}
\end{align}
Using the variational condition $\delta \tilde{A}_{1}=0$ we can compute the correction terms $c_{n}$. For the
linear correction term (n=1) we obtain

\begin{eqnarray}
\label{da12}
\delta A_{1}(x)+\int dy  c_{1}(x,y)\delta \Omega_{1}(y)=0 \,.
\end{eqnarray}
Combining Eqs.~(\ref{4.67}), (\ref{4.71}) and (\ref{da12}) we have 

\begin{eqnarray}
\label{4.75}
 c_{1}(x,y) = \frac{1}{e^{2}(\alpha -1)}\partial_{1}\delta(x-y)\,. 
\end{eqnarray}
For the quadratic term we have $c_2=0$. Then, all the correction terms $c_{n}$ are null for $n\geq 2$. Using Eq.~\eqref{4.75} in
\eqref{37} we can obtain the GU variable $\tilde{A}_{1}(x)$ 

\begin{eqnarray}
    \tilde{A}_{1}(x) = A_{1}(x)-\frac{1}{e^{2}(\alpha -1)}\partial_{1}\Omega_{1}(x) = A_{1}(x)-\frac{1}{e^{2}(\alpha -1)}\partial_{1}\pi_0(x)\,.
\end{eqnarray}
Repeating this same iterative process we can derive the others GU variables which are

\begin{align}
&\tilde{A}_{0}=A_{0}  \,,  \label{e1} \\
&\tilde{\varphi}=\varphi+\frac{1}{e(\alpha -1)}\,\pi_0\,, \label{e2} \\
&\tilde{\pi}_{1}=\pi_{1}+\frac{1}{(\alpha -1)}\,\pi_0\,, \label{e3} \\
&\tilde{\pi}_{\varphi}=\pi_{\varphi}+\frac{1}{e(\alpha -1)}\,\partial_{1}\pi_0\,. \label{e4}
\end{align}
Replacing $A_{0}, A_{1}, \varphi, \pi_{1}$ and $ \pi_{\varphi}$ by $\tilde{A}_{0},\tilde{A}_{1},\tilde{\varphi},\tilde{\pi}_{1}$ and $\tilde{\pi}_{\varphi}$ in the canonical Hamiltonian, Eq.~\eqref{6}, thus we can derive the gauge invariant Hamiltonian $\tilde{H}^{*}_{c}$ 

\begin{align}
\tilde{H}_{c}^{*}= H_{c} + \int dx \bigg[ \frac{[\pi_{1}+ (\alpha -1)\partial_{1}A_{1}]}{(\alpha -1)}\pi_{0} + \frac{(\partial_{1} \pi_{0})^{2}}{2e^{2}(\alpha -1)}+ \frac{\pi^{2}_{0}}{2(\alpha-1)^{2}}  \bigg]\,,\label{4.109}
\end{align}
which by construction satisfies $\left \{ \chi, \Tilde{H}^{*}_{c} \right \}=0$.

The Poisson brackets between the GU variables are

\begin{align}
\left \{ \tilde{A}_{0}(x),\tilde{\pi}_{1}(y) \right \}&=\frac{1}{(\alpha - 1)}\delta(x-y) = \left \{ A_{0}(x),\pi_{1}(y) \right \}_{D}\,,\\
\left \{ \tilde{A}_{0}(x),\tilde{\varphi}(y) \right \}&=\frac{1}{e(\alpha - 1)} \delta(x-y)=\left \{ A_{0}(x),\varphi(y) \right \}_{D}\,,
\end{align}
\begin{align}
\left \{ \tilde{A}_{0}(x),\tilde{\pi}_{\varphi}(y) \right \}&=-\frac{1}{e(\alpha -1)}\partial_{1}\delta(x-y)=\left \{ A_{0}(x),\pi_{\varphi}(y) \right \}_{D}\,,\\
\left \{ \tilde{A}_{0}(x),\tilde{A}_{1}(y) \right \}&= \frac{1}{e^{2}(\alpha -1)}\partial_{1}\delta(x-y) = \left \{ A_{0}(x),A_{1}(y) \right \}_{D}\,,\\
\left \{ \tilde{A}_{1}(x),\tilde{\pi}_{1}(y) \right \}&=-\delta(x-y) =\left \{ A_{1}(x),\pi_{1}(y) \right \}_{D}\,,\\
\left \{ \tilde{\varphi}(x),\tilde{\pi}_{\varphi}(y) \right \}&= \delta(x-y) =\left \{ \varphi(x),\pi_{\varphi}(y) \right \}_{D}\,.
\end{align}
As we have obtained in Case I, the Poisson brackets between the GU variables reduce to the original Dirac brackets between the original
second-class variables.

\section{A comparison with the Batalin-Fradkin-Tyutin (BFT) method}

In this section, we will make a brief comparison of our results with the gauge invariant variables of the BCSM 
obtained with the help of the newly improved BFT formalism. For more details about this procedure see reference
~\cite{A_22}. This method, in stark contrast to the improved GU 
formalism, makes an enlargement of the phase space with extra variables while extracting a gauge theory from a second-class constrained system.

Initially we will consider the original second-class function $G=(A^{\mu}, \pi_{\mu}, \varphi, \pi_{\varphi})$ in the BCSM. Then, the new 
first-class function $\Tilde{G}=(\Tilde{A}^{\mu}, \Tilde{\pi}_{\mu}, \Tilde{\varphi}, \Tilde{\pi}_{\varphi})$ in the extended phase space 
can be expressed as
 
 \begin{align}
 \label{bf1}
\Tilde{G}(A^{\nu}, \pi_{\nu}, \varphi, \pi_{\varphi};\Phi^{j})=G +\sum_{n=1}^{\infty}\Tilde{G}^{(n)}, \qquad \Tilde{G}^{(n)}\sim (\Phi^{j})^{n}
 \end{align}
 where $\Phi^{j}$ are the two extra variables. The function $\Tilde{G}$ satisfies the following boundary conditions: 
 
 \begin{align}
 \Tilde{G}|_{\Phi^{i}=0}=G \,.
 \end{align}
The first order correction terms in $\Phi^{j}$ are given by

\begin{align}
 \Tilde{A}^{\mu (1)} &=\left ( \frac{1}{e\sqrt{\alpha - 1}}\Phi^{2}, \frac{1}{e\sqrt{\alpha - 1}}\partial^{1}\Phi^{1} \right ) \,,\\
\tilde{\pi}_\mu^{(1)}&=\left ( e\sqrt{\alpha - 1} \Phi^{1}, -\frac{e}{\sqrt{\alpha - 1}}\Phi^{1} \right ) \,, \\  
\Tilde{\varphi}^{(1)}&= -\frac{1}{\sqrt{\alpha - 1}}\Phi^{1}, \, \\
\Tilde{\pi}_{\varphi}^{(1)}&= \frac{1}{\sqrt{\alpha - 1}}\partial^{1}\Phi^{1} \,. 
 \end{align}


\noindent In the sequence we have that the higher order correction terms are null. Then, the BFT gauge invariant variables can be written as

\begin{align}
\label{bft21}
\Tilde{A}^{\mu}&=A^{\mu}+\Tilde{A}^{\mu (1)}=\left ( A^{0} + \frac{1}{e\sqrt{\alpha -1}}\Phi^{2}, A^{1}+\frac{1}{e\sqrt{\alpha -1}} \partial^{1}\Phi^{1} \right ) \,,\\
\label{bft22}
\Tilde{\pi}^{\mu}&=\pi^{\mu}+\Tilde{\pi}^{\mu (1)}=\left ( \pi^{0} + e\sqrt{\alpha -1}\Phi^{1}, \pi^{1}+\frac{e}{\sqrt{\alpha -1}} \Phi^{1} \right ) \,,\\
\label{bft23}
\Tilde{\phi}&=\phi+\Tilde{\phi}^{(1)}= \phi-\frac{1}{\sqrt{\alpha -1}} \Phi^{1} \,,\\
\label{bft24}
\Tilde{\pi}_{\phi}&=\pi_{\phi}+\Tilde{\pi}^{(1)}_{\phi}= \pi_{\phi}+\frac{1}{\sqrt{\alpha -1}}\partial^{1} \Phi^{1} \,.
\end{align}
 Using an elegant property~\cite{A_22}

 \begin{align}
\Tilde{H}(A_{\mu}, \pi_{\nu}, \varphi, \pi_{\varphi};\Phi^{i})=H_{c}(\Tilde{A}_{\mu}, \Tilde{\pi}_{\nu}, \Tilde{\varphi}, \Tilde{\pi}_{\varphi}) \,,
 \end{align}
we have
 
\begin{align}
\label{bft3}
\Tilde{H}(A^{\mu}, \pi_{\nu}, \varphi, \pi_{\varphi};\Phi^{i})&=H_{c}(A^{\mu}, \pi_{\nu}, \varphi, \pi_{\varphi}) +\int dx\bigg [ \frac{1}{2}(\partial_{1}\Phi^{1})^{2} +\frac{e^{2}}{2(\alpha -1)}(\Phi^{1})^{2}+\frac{1}{2}(\Phi^{2})^{2} \nonumber \\
&+\frac{1}{e\sqrt{\alpha - 1}}\left [ e^{2}\pi^{1}-e^{2}(\alpha-1)\partial_{1}A_{1}  \right ]\Phi^{1} - \frac{1}{e\sqrt{\alpha - 1}}\Tilde{\Omega}_{2}\Phi^{2} \bigg ].
\end{align}
If we make the identification $\Phi^{1}=-\frac{\Omega_{1}}{e\sqrt{\alpha-1}}$ and $\Phi^{2}=0$ in the Eqs.~(\ref{bft21}), (\ref{bft22}),
(\ref{bft23}), (\ref{bft24}) and (\ref{bft3}) we can observe that the gauge invariant variables and the gauge invariant Hamiltonian, 
both obtained with the help of the BFT formalism, coincide 
with the GU variables, Eqs.~(\ref{e1}), (\ref{e2}), (\ref{e3}) and (\ref{e4}) and the GU Hamiltonian, Eq.~(\ref{4.109}). This result can indicate the consistency of our formalism.
 
\section{Final Remarks}

In this work we have converted the BCSM with $ \alpha > 1$ into a first-class system by using the improved GU formalism.
Initially, the canonical structure of this model was analyzed by the Dirac formalism for constrained systems where it was observed that for $\alpha > 1$ the system presents two second-class constraints. Subsequently, the improved GU formalism was applied which has provided two distinct gauge theories. In case I, the rescaled  primary constraint, Eq.~(\ref{g1s}), was chosen to be the gauge symmetry generator and the corresponding  GU variables are derived. In case II, the rescaled secondary constraint, Eq.~(\ref{rsecond}), was chosen to be the gauge symmetry generator and we also obtain the respective GU variables. Furthermore, we have shown precisely
that the improved GU procedure leads to the same Poisson brackets
for both cases I and II. This result can also indicate the consistency of our formalism. Finally, we have compared our results obtained by using the improved GU formalism with those obtained by using the improved BFT method.
 We have observed that the gauge invariant variables and the gauge invariant Hamiltonian, both derived with the aid of the improved BFT formalism, coincide with the GU variables and the GU Hamiltonian of case II when we take particular values for the Wess-Zumino fields. Here we can mention that the improved GU formalism leads to less extensive calculations compared to the usual GU method which the canonical Hamiltonian is directly converted into a gauge invariant Hamiltonian. Therefore, as we have observed throughout the work, once the GU variables are determined then the corresponding gauge theory is obtained in a very straightforward way.

\label{FR}

\section{Acknowledgments}
The CAPES (Coordenação de Aperfeiçoamento de Pessoal de Nível Superior), FAPEMIG (Fundação de Amparo à Pesquisa do Estado de Minas Gerais) and UFJF (Universidade Federal de Juiz de Fora) are ackknowledged for financial support. Jorge Ananias Neto thanks CNPq (Conselho Nacional de Desenvolvimento Cient\'ifico e Tecnol\'ogico), Brazilian scientific support federal agency, for partial financial support, CNPq-PQ, Grant number 307153/2020-7.

\end{document}